# Applying Module-less Synthesis on Cyberphysical Digital Microfluidic Biochip Ensuring Error Detection and Routing Performance Optimization

Sarit Chakraborty
Computer Science & Technology
Indian Institute of Engineering Science & Technology
Shibpur, Howrah, India
E-mail: sarit.rs2014@cs.iiests.ac.in

Susanta Chakraborty
Computer Science & Technology
Indian Institute of Engineering Science & Technology
Shibpur, Howrah, India
E-mail: sc@cs.iiests.ac.in

***Abstract*— Digital Microfluidic Biochips (DMFB's) consist of Two-Dimensional (2-D) microarrays that are integrated with different healthcare related cyberphysical systems and expected to be used extensively in near future. Thus, faster and error-free synthesis techniques need to be implemented on such chips. Various Bio-protocols are performed based on different mixing modules present on the chip until now. In this work, the concept of such dedicated virtual modules has been eliminated and a novel Module-Less-Synthesis (MLS) method is proposed for accomplishing bioassays for cyberphysical DMFB's. However, path congestion problem and operational errors are inevitable in MLS approach.**

**We have identified various shift-patterns (movements) to accomplish entire mixing in lesser time compared to earlier synthesis methods and computed the percentage of mixing accomplishment for each directional-shift of the mixer-droplet. A novel error detection mechanism and routing optimization for MLS method is also proposed using satisfiability solver. Finally, the path congestion and washing problem in MLS is addressed by tweaking the MLS approach and a new modified-MLS method is proposed. Experimental results on real-life bioassay like PCR, IVD and on synthetic benchmarks (hard test benches) show improved synthesis performance with bioassay completion time ($T_{max}$) significantly reduced from earlier synthesis approaches. The proposed error detection mechanism ensures efficiency of the chip with a recovery rate of more than 95% for normal chip sizes and using SAT solver we are able to find recovery shift-movements even on smaller chips by avoiding the deadlocks.**

## I. INTRODUCTION

**D**igital **M**icrofluidics is the new paradigm used for monitoring and prognostic applications in the field of medical, pharmaceutical and environmental sciences [1]. It can manipulate liquids as discrete volume ($10^{-9}$ or $10^{-12}$ ltr.) of droplets on a 2-D array of electrodes [2]. Initially microfluidic biochips [1] were based on micro-pumps and microvalves, and continuous liquid flow through fabricated microchannels present on such chips. Presently Droplet-based "digital" microfluidics technology [2] are merged with software-controlled, physical-aware (cyberphysical) DMFBs which can be extensively used for Point-of-Care (PoC) tests and treatments where adequate laboratory facilities are missing, especially in developing countries [6].

In the first work on cyberphysical DMFB [5], a rollback strategy is used to handle error recovery but all recovery operations are carried out in a stand-alone manner. All other ongoing bioassay-related fluidic operations are interrupted when an error is detected. The error-recovery approach presented in [5] cannot handle situations when multiple errors occur during a bioassay synthesis. Another methodology was given [6], where the control software re-computed the schedule of fluidic operations, module placement, and droplet pathways depending on the sensor feedback. Despite its novelty and advantages, the given error-recovery method [6] suffers from several shortcomings. The online re-synthesis [6] involves software-based dynamic regeneration of electrode actuation sequences. Such a resynthesize step leads to increased bioassay response time and all fluid handling operations are interrupted when an error occurs. A dynamic online decision-making methodology is mentioned [7] for the execution of fluidic operations and droplet routing in response to the detection unit. Here the progress of the assay solely depends on the feedback from the sensing system associated with the biochip. A dictionary based error recovery technique that has the capability of handling at most two errors are presented in [8]. The dictionary is pre-computed for the entire bioassay and is stored in a microcontroller and assay operations take place accordingly. However, the methodology [8] did not discuss the presence of any routing information in the dictionary needed to route the droplet from one mixer to another or to waste reservoir in case of occurrence of errors. Moreover, the two compaction methods given [8] offer very minimum compaction ratios.

Literature available presently shows only Module-based synthesis techniques are used till date on Cyberphysical DMFB platform [4], [7], [12], [19]. This, in turn, increases the completion time of the bioassays because a limited number of "mixing-modules" are available at any point of time on the chip. Moreover, module-based synthesis methodologies decrease the chip utilization factor [18] as huge numbers of guard cells are occupied during the module operations to pad the mixer droplets from cross-contamination with other existing droplets on the chip. Maftei et al. [22] first proposed the concept of Routing based synthesis. Mixing time was improved by avoiding flow reversibility and eliminating static pivot points present in different mixing modules. However, no appropriate computation [22] of mixing completion time was addressed for consecutive linear movements of the mixer-droplet. Thus it cannot be suitably applied on cyberphysical DMF chips. In addition, a simplistic assumption of straight flow of the mixer droplet is assumed [22] for faster mixing completion but no discussion about the limitation of linear flow is given. In [23], congestion avoidance and washing mechanisms are mentioned for routing-based synthesis technique. The chip operating frequency assumed to be100Hz, which produces a switching time of 10 ms [23]. Nevertheless, most of the applications on DMF chips involve fluidic operations on viscous analytes or medium, making it very difficult to apply such fast switching. In addition, it cannot be applied to cyberphysical chips where real-time error detection and error control are two important aspects and need to be addressed separately.

In this work, we present a module-less synthesis methodology for cyberphysical DMFB (CP-DMFB) that ensures better chip utilization by omitting the virtual modules on the chip and extra electrodes as guard cells. The droplets are free to move on the 2-D biochip by different types of shift-movements. Mixing is done through diffusion rather occupying a dedicated region on the chip and assay completion times are reduced by our proposed methodology as well as completion time uncertainties are also decreased. Thus it satisfies cyberphysical DMFB's operational criteria of faster synthesis time and error handling capability by adopting definite paths for each of the mixing stages. However, in MLS approach path congestion complexities and errors are inevitable due to the integration of cyberphysical paradigm. Our technique efficiently handles congestion and reduce washing overhead by tweaking earlier MLS approach [18] and proposing the modified-MLS (MMLS) method. An efficient error detection mechanism is also proposed to achieve faster assay completion ensuring the highest possible accuracy of the assay results.

The key contributions of this paper are as follows.

1. Section II discusses the preliminary concepts of basic construction, droplet operations and synthesis steps on a CP-DMFB.
2. The MLS problem formulation is shown in Section III derived from the existing mixing modules available in DMFB domain Library.
3. A novel Module-Less Synthesis (MLS) approach is proposed for cyberphysical DMFB in Section IV. Mixing completion percentage for each directional-shifts are derived in Sub-Section IV.(A) and MLS algorithms are presented in Sub-Section IV.(B)
4. To avoid congestion in MLS approach and to address washing optimization, a new Modified-MLS method is proposed in Sub-Section IV.(C) and use of SMT-Solver model for deadlock avoidance is presented in sub-section IV.(D).
5. Error recovery problem for MLS approach on cyberphysical DMFB and an efficient error detection strategy is addressed in Section V.
6. In Section VI a detailed illustrative example of PCR synthesis is given by our MLS approach and simulation results are presented for two representative bioassay as well as for synthetic hard test benches in Section VII.
7. Finally Section VIII concludes the MLS approach for cyberphysical DMFB's.

## II. PRELIMINARIES

### A. *Basic Construction and Operations of DMFB cell*

The micro-droplets may consist of biomedical samples like blood, serum, urine or saliva and the filler medium (usually silicone oil) are sandwiched between two parallel glass plates as shown in Fig. 1. The bottom plate consists of a series of individually controllable electrodes and a single top plate is used as ground electrode. The droplet movement is achieved by the principle of 'Electro-Wetting-on-Dielectric' (EWOD) [4]. The cyberphysical DMFB instruction set includes droplet transport in four directions i.e. UP, DOWN, RIGHT and LEFT as shown in Fig. 2a. Splitting a droplet into two unit volume droplets and merging of two droplets into one is shown in Fig. 2b. Similarly, mixing, storage (incubation) and detection of the microdroplets is also inevitable operations for accomplishing any real-life assay on the cyberphysical chip [19]. Apart from that, external devices such as heaters, photodetectors [5], [20] capacitance sensors, impedance sensors [21] are used to offer additional functionalities.

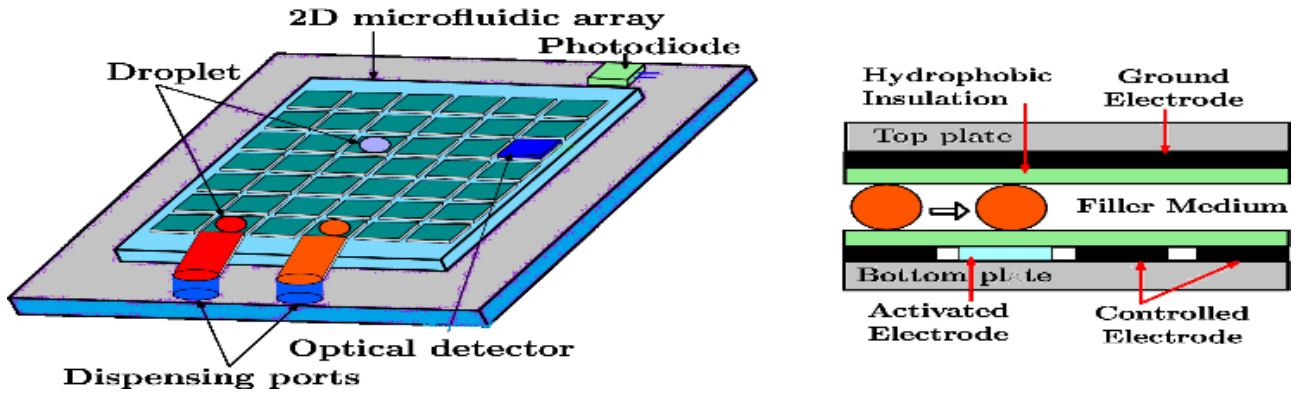

**Fig. 1**: Basic construction of a DMFB and droplet movement

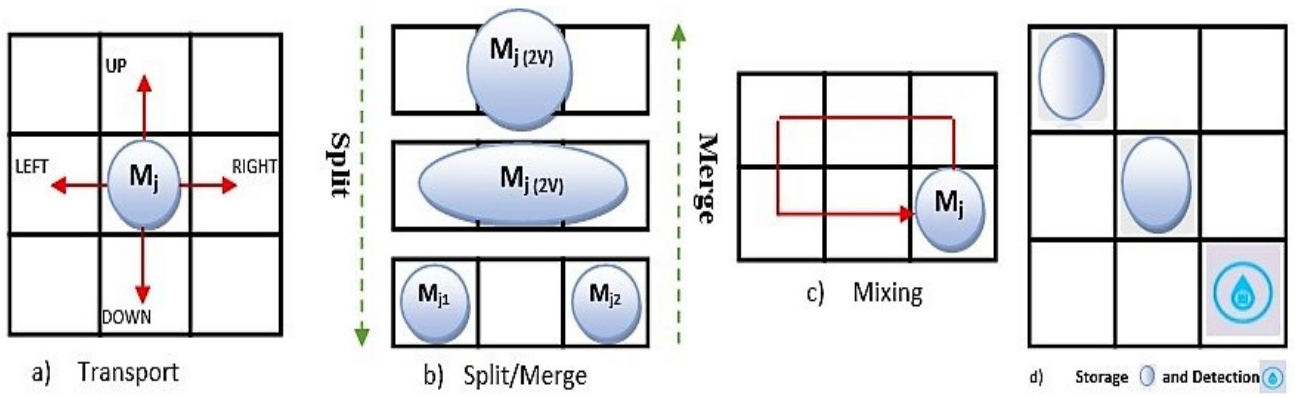

**Fig. 2:** Basic set of droplet operations performed on a DMFB.

### B. *Synthesis flow for DMFB*

The synthesis on a CP-DMFB starts with a sequence graph (G) which is derived for a specific bioassay to be performed and it continues through several stages such as binding, scheduling, placement and routing (Fig. 3.). A module library is declared which holds the realization information such as mixing, detection etc. with their respective grid-size and timing requirements [9]. The modules are bind with nodes of the graph (G) and scheduling determines the order of operations based on such binding. Finally, the placement of such modules on the chip is determined and routing starts.

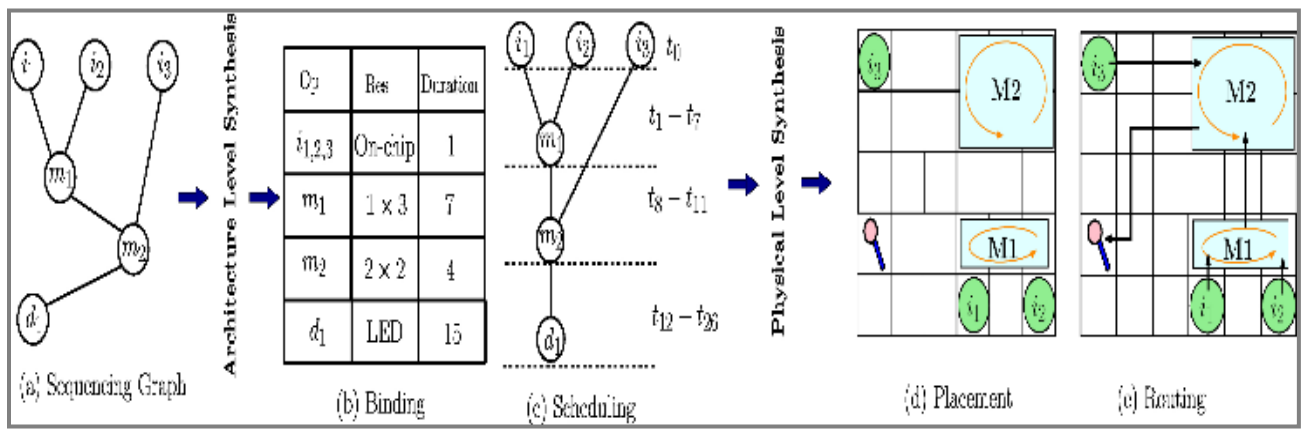

**Fig. 3**: Synthesis steps for typical bio-operations on DMFB

### C. *Cyberphysical Digital Microfluidic Biochips*

Cyberphysical inclusion with DMF-chips denotes integration of sensing system, a feedback network and a control software along with the conventional DMFB [7]. Such software-based reconfigurability option for DMFB has motivated the research on various aspects of automated chip design, chip applications and chip optimization in one level up [19]. This, in turn, also improves the performance and provide the scope of efficient error-recovery for the biochips [20]. The control software sends a control signal to the microfluidic biochip, and the on-chip sensing system monitors the outcome of fluidic operations. The outcomes are then compared with the expected reference values as shown in Fig. 4. If an error has occurred, the control software receives a "repeat request", and the corresponding operation in which the error has occurred can be re-executed. Thus, the intermediate results can determine the further steps of the bioassay in real-time which can obviously produce more precise and accurate outcome of the chip as explicitly needed in Point-of-Care diagnostics.

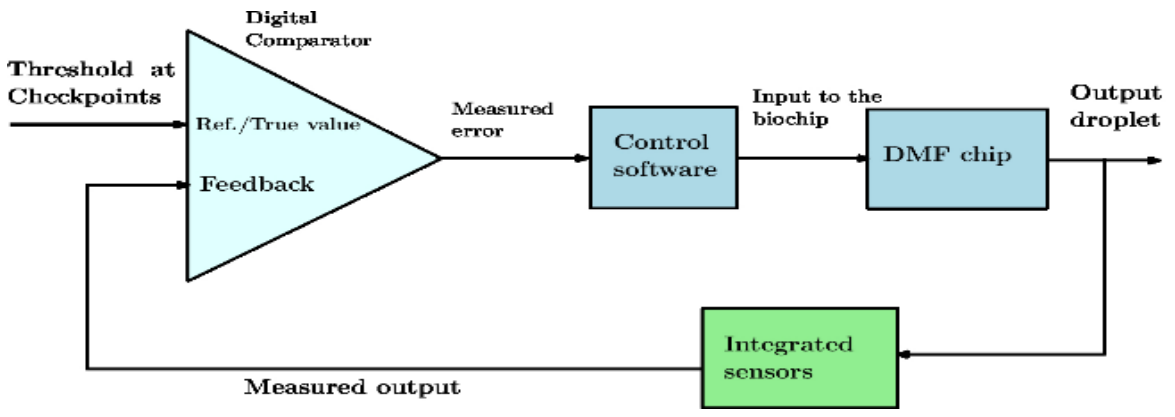

**Fig. 4.** Schematic of a typical cyberphysical DMFB system

A typical cyberphysical system has various components as follows:

*a)* Interface Sensing System, which is used to sense various physical parameters of the micro/nano-droplets e.g. mass, volume, concentration, color etc. with the help of an integrated Optical Detector [15], [16] or Charge-Coupled Device (CCD) based camera [17]. The camera snapshot is then compared with the reference value via a software control system integrated with cyberphysical DMFB [20].

*b)* Physical-aware Software, which has the capability of using sensor data at intermediate checkpoints to detect errors and act consequently to minimize the impact of errors.

*c)* Interfaces between Biochip and Control Software establishes the cyberphysical coupling between the control software and the hardware of the microfluidic platform [20].

## III. Problem Formulation

The synthesis methodologies present in the literature of cyberphysical DMFB use module based synthesis. A new approach is shown in [7], where smaller mixing modules are used at initial stages, merged in bigger modules, as synthesis progresses and ultimately entire chip becomes a single mixing module. However, no routing information about the intermediate stages is discussed. We have considered the module library given in [10] as a standard for DMFB platform, where timing requirements of different fluidic operations are provided. Fig. 5 shows different mixing modules along with their respective padding cells typically used in assay synthesis on a CP-DMFB. In addition, the mixing completion time in respect of ascending module sizes and respective number of padding cells occupied during the module operations are given in Table 1. From Table 1, it is clear that no correlation exists between the module sizes (Active Mixing Region) and mixing completion time. For a 2 x 2 and 1 x 4 module consisting of same area on the chip, mixing time varies more than 100%. The fastest mixing module is 2x4, where the mixing completion time is 2.9 sec. But, this 2x4 module actually blocks 24 cells (electrodes) when in use, though only 8 electrodes are participating during the entire mixing operation and rest of the 16 cells work as padding cells (Table 1) to avoid cross-contamination with other droplets as shown in Fig. 5(c). Moreover, only 1 electrode is used at a particular time instant (t) where t = 1 time-step and it is in the range of approximately 0.05 sec $\leq t \leq$ 0.125 sec depending on the operating frequency (f) of the chip which generally varies between 20Hz to 8Hz for a CP-DMFB. Thus, in module-based synthesis methodology, both hardware cost (more than 66% module area are wasted during the module operations) and bioassay completion time are large enough and hence unsuitable for cyberphysical DMFB requirements.

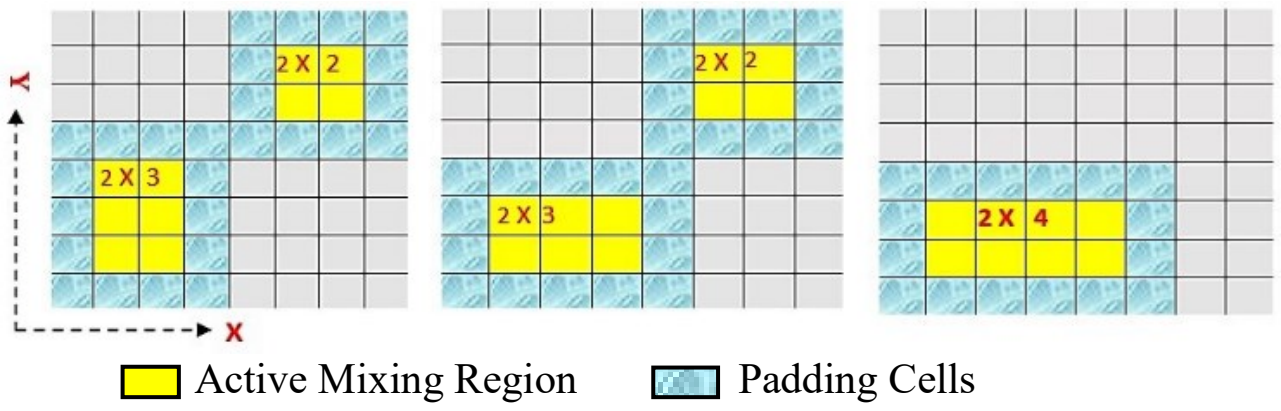

**Fig. 5.** Placement of Mixing Modules along X-Y axis

**Table 1:** Module Library of DMFB

| *Operation* | *Module Area (# of Active Cells)* | *Blocked Area (# of Padding Cells)* | *Mixing completion Time (sec)* | *Unused Area ( in Percentage)* |
|---|---|---|---|---|
| Mixing | 1 X 4 | 14 | 4.6 | 77.7 |
| Mixing | 2 X 2 | 12 | 9.95 | 75.0 |
| Mixing | 2 X 3 | 14 | 6.1 | 70.0 |
| Mixing | 2 X 4 | 16 | 2.9 | 66.6 |
| Dispensing | -- | | 2 | |
| Detection | 1 X 1 | | 30 | |

We have formulated the problem considering the PCR bioassay [11] and a module based chip of size 8X9 cells shown in Fig. 6. We have placed 4 mixing-modules on the chip to accommodate 1st layer of PCR simultaneously and omitted the 2x2 module which having the longest time requirement (Table 1). Now if $M_1$, $M_2$ operations are performed on 1x4 mixer-units and $M_3$, $M_4$ runs on 2x4 and 2x3 mixer units respectively, then $M_1$, $M_2$ operations will be completed in 4.6 sec and $M_3$, $M_4$ finishes respectively in 2.9 secs and in 6.1 secs. As $M_5$ is the mixing of $M_1$ and $M_2$, thus $M_5$ can directly start after 4.6sec whereas $M_6$ cannot be started before the completion of $M_4$ (6.1 sec). In next layer, if $M_5$ and $M_6$ run on 2x4 and 1x4 mixing modules (the fastest possible mixing units available on the chip), their completion time will be 7.5 sec and 10.6 sec respectively. Thus, it is obvious that $M_7$ will be completed only after 13.6 (≈14) secs. Thus, the completion time of the PCR assay requires 14 secs in module based cyberphysical DMFB ignoring the time required for other operations like split, merge, and detection of the droplet. Moreover, to fit the 1st layer of PCR sequence graph (G) with 4 parallel mixing operations (nodes) as per above module configurations, no other way the modules can be placed on an 8x9 (72 cells) chip size to avail more free cells for other operations. Out of total 72 electrodes (cells), only 9 cells are available for other operations like detection, dispensing etc.

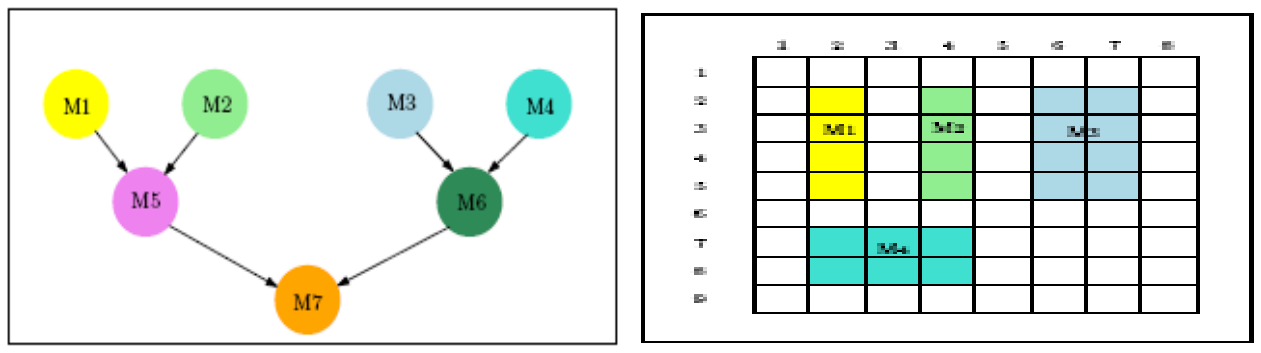

(a) Sequence graph (G) for PCR b) Placement of different Modules

**Fig. 6**: Module-Based Synthesis on DMFB

Thus, module-based chip demands bigger chip size and ultimately results in wastage of spatiotemporal resources on the chip. It also requires a huge number of activation pins, which essentially draws more power and consequently chip degradation is inevitable.

In module-based mixing, we can detect the error only after the mixing process fully completed within the virtual modules. Thus, if any error occurs the entire mixing time is wasted and we need to restart the mixing from the same initial state in the same error-prone module or in a different module available on the chip at that point of time. Thus, module-based synthesis approach involved more synthesis time and more overhead for the recovery process, if any error occurs.

## IV. Proposed Synthesis Method

We have proposed a new synthesis technique eliminating the usage of virtual mixing modules on the DMFB chip. The usual operations like mixing, splitting, merging, routing and detection is generally performed on a DMFB by basic steps of droplet movement through different electrode activation sequences. This work primarily focuses on the synthesis of various bioassays by adopting definite path for the mixer droplets. Thus, it satisfies the demand of minimizing assay completion time and error-free results for a cyberphysical DMFB. Diffusion-based mixing is considered on the basic framework of DMFB as mentioned in Paik et al. [10]. We have proposed different directional shifts (movement) of the mixer-droplet and for each shift-movement; the percentage of mixing accomplishment is computed. In addition, an efficient error detection approach is also formulated.

### *A. Computation of Mixing Completion for Various Directional-Shifts*

This section deals with the mixing completion percentage for various directional-shifts of the droplet. We derive the time based on inductive proof by taking the experimental results of [10]. In [10] authors have performed laboratory experiments on 1x4, 2x2, 2x3 and 2x4 mixing modules and a module library is prepared (Table 1). Mixing of droplets is to be carried out in various mixing modules following a specific pattern of shifts (directional-movement) of the droplets as shown in Fig. 7. Thus, it is seen the various shifts a mixer-droplet can adopt are broadly categorized into three types. These are $0^\circ$-shifts (zero-degree shift-movements), $90^\circ$-shifts (ninety-degree shift-movements) and $180^\circ$-shifts. We have further divided the $0^\circ$ shift-movement into three types. $Z_1$-shift which is $0^\circ$-shift along one cell position as shown in 2x3 mixture, $Z_2$ -shift along two consecutive cells as shown in 2x4 and 1x4 mixing modules in Fig. 7(a) and 7(b) respectively. The $Z_3$-shift (three consecutive straight runs of the droplet) can be derived from Fig. 8. For the sake of clarity, the following naming conventions are given in Table 2 and will be used in later part of this paper.

**Table 2:** Various directional Shift-Names

| ***SHIFT TYPE*** | ***SHIFT NAME*** | ***Sub Classification*** |
|---|---|---|
| **$90^\circ$** | X-shift | **NULL** |
| **$180^\circ$** | Y-shift | **NULL** |
| **$0^\circ$** | Z-shift | **$Z_1$, $Z_2$, and $^{**}Z_3$** |

The standard operating frequency (f) of the cyberphysical chip is chosen as 16 Hz. Thus the time (t) of moving a droplet from one electrode to any of its four neighboring electrodes is

$$t = (1 / f) = 1/ 16\text{Hz} = 0.0625 \text{ Sec} \quad (1)$$

Now, the calculation of percentage completion of mixing (dilution) for each type of shift-movement is given as follows:

First, we have considered the 2x2 mixing module (Fig. 7d) and calculate mixing completion percentage for each $90^\circ$ shift (X-shift). From Fig. 7(d) it can be seen a 2x2 module can have only 4 different X-shifts in a cycle and total time for mixing completion in a 2X2 mixer is 9.95sec. So number of shifts required = 9.95 / 0.0625 = 160 steps (approx.).

Now considering 160 steps of X-shift ≡ 100% mixing completion. So 1 step of X-shift can accomplish (100 ÷ 160) % = 0.625% of mixing completion. After establishing mixing completion percentage for each X-shift, we consider the next bigger mixing module of size 2x3 (Fig. 7c), where each $Z_1$-shift are preceded and followed by two consecutive X-shifts. The mixing completion percentage by each $Z_1$-shift is calculated below.

Total time for mixing completion in 2X3 mixer = 6.1sec. So, number of shifts required = 6.1 / 0.0625 = 96 steps (approx.). Thus 64 steps of X-shift + 32 steps of $Z_1$-shift ≡ 100% mixing. i.e. 64X + 32 $Z_1$ = 100%. So, 32 steps of $Z_1$-shift = (100 – 64*0.0625) % of mixing = 60% mixing. So 1 step of $Z_1$-shift can accomplish (60 /32) % = 1.875% of mixing.
Similarly considering 2x4 mixing module, the mixing completion for each $Z_2$-shift (two consecutive straight-run of the mixer droplet) is 5% and finally, each $180^{\circ}$ shift (Y-shift) accomplishes -2.5% of mixing. The negative mixing for Y-shift is explained by the unfolding of patterns inside the droplet, i.e., two droplets tend to separate when moved backward ($180^0$-shift). Thus, it is seen consecutive linear shifts produce faster mixing. But no limitations on a maximum number of linear shift is derived till date.

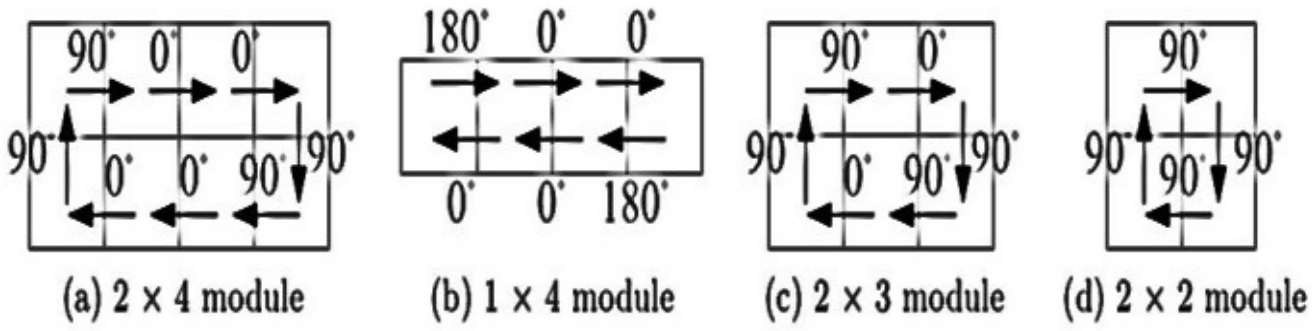

**Fig. 7**: Various Mixing Modules as per the available Module Library

***Lemma 1:*** **A maximum of 3 consecutive straight/linear movement of the droplet is possible in MLS method.**

***Proof.*** Using the module library statistics for DMFB as shown in Table 1**,** it can be seen that 2x4 is the maximum module size available in the literature [10].We have derived the mixing completion percentage for X, Y, $Z_1$, $Z_2$-shifts. Now having the said values we have applied Lagrange's Interpolation formula to find the probable completion time of mixing (dilution) for next bigger (2x5) mixing module, where $x_1, x_2, x_3$ denotes module length along the X-axis and $y_1, y_2, y_3$ denotes mixing time for each module. Now, putting $x = 5$, we get $f(5) = 1$. Thus, probable mixing completion time for a 2x5 module is 1 sec.

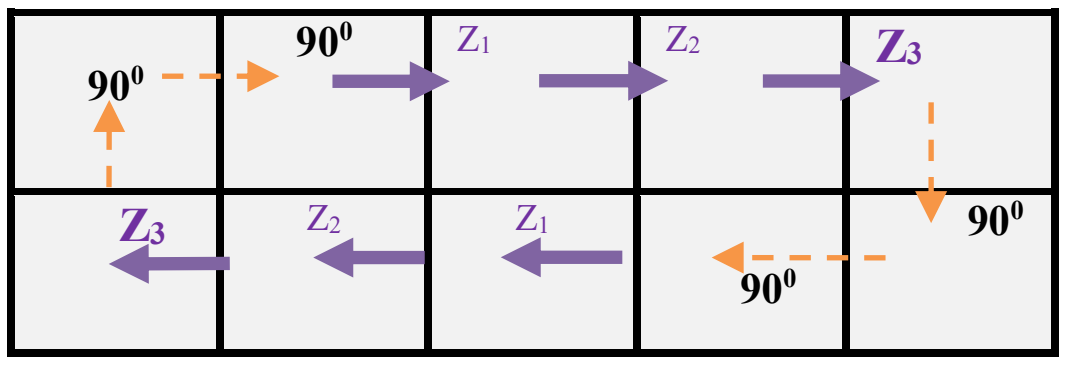

**Fig. 8**: 2X5 Mixing Module

From Fig. 8, it can be seen the shift-pattern in a 2x5 mixer module, where each mixing cycle consists of $Z_3$-shift (three consecutive $0^0$-shifts) followed by two X-shifts ($90^0$-shift). If we calculate the mixing completion amount for each $Z_3$-shift, we have found it is more than $Z_2$-shift and approximately 15% mixing completion is achieved for each $Z_3$-shift. It essentially establishes the assumption of more number of one directional straight-movement (consecutive Z-shifts) of the mixer droplet along the same coordinate axis can accomplish faster mixing than combining horizontal and vertical movement one after another. Thus a 1x4 module can finish much faster mixing than a 2x2 module, though both having the same area. Thus, mixing pattern (directional-shift-movements) can essentially reduce the mixing time a lot as well as overall synthesis time of the assays. However, if we fix the module size (2xN) as in module-based approach [10], we essentially restrict the shift-pattern a mixer-droplet may have. So using Lagrange's formula shown in equation (2), we extrapolate the timing requirements for next bigger module (2xN, where N=6) to derive next higher consecutive liner movements of the droplet but arrived at a negative (–ve) mixing completion time. A module having negative mixing-completion time is an infeasible condition. Hence maximum module size possible on a DMFB is 2 x N cells where 2≤N≤5.

$$f(x) = \frac{(x-x_2)(x-x_3)}{(x_1-x_2)(x_1-x_3)} * y_1 + \frac{(x-x_1)(x-x_3)}{(x_2-x_1)(x_2-x_3)} * y_2 + \frac{(x-x_1)(x-x_2)}{(x_3-x_1)(x_3-x_2)} * y_3 \qquad (2)$$

$$f(x) = \frac{(x-3)(x-4)}{(2-3)(2-4)} * 10 + \frac{(x-2)(x-4)}{(3-2)(3-4)} * 6 + \frac{(x-2)(x-3)}{(4-2)(4-3)} * 3$$

$$f(x) = 5 * (x-3)(x-4) - 6 * (x-2)(x-4) + 1.5 * (x-2)(x-3)$$

$$f(x) = 5x^2 - 35x + 60 - 6x^2 + 36x - 48 + 1.5x^2 - 7.5x + 9$$

$$f(x) = 0.5x^2 - 6.5x + 21$$

Thus from Lemma 1, we have concluded, a maximum number of three consecutive Z-shift (straight-run of the droplet) is possible which we named as $Z_3$ and shown in Fig. 8. As it is infeasible to have a 2x6 or bigger module, after each $Z_3$-shift the droplet has to take a mandatory X-shift. Various possible shift-movements with their corresponding percentage completion of mixing are tabulated in Table 3.

**Table 3:** Mixing-Completion for Various Droplet-Shifts

| TYPE OF SHIFT's | Mixing- completion in single shift movement ( in % ) |
|---|---|
| X ($90^0$) | 0.625 |
| Y($180^0$) | -2.5 |
| $Z_1$ ($0^0_1$) | 1.875 |
| $Z_2$ ($0^0_2$) | 5.0 |
| $Z_3$ ($0^0_3$) | 15.0 |

By following the proposed shift-movements (X, Y, $Z_1$, $Z_2$ and $Z_3$); the mixer-droplet can now move on the entire chip as per the availability of chip space and maintaining the DMFB routing constraints. Thus, it can definitely eliminate the requirement of fixed mixing-modules on the chip as well as it reduces the overall mixing (synthesis) time. The minimum time-steps (t) required for mixing completion can be formulated as below.

***Lemma 2:*** **For 16 Hz operating frequency DMF chip, the minimum number of time-steps required for full mixing completion (100%) is 27 time-steps.**

***Proof.*** As per proposed MLS method, the maximum amount of mixing completion can be achieved by the $Z_3$-shift (15%), which consists of three consecutive straight runs of the mixer droplet. Also after having a $Z_3$-shift, it is mandatory to take an X-shift **(**$90^0$ -move**)** before acquiring another linear movement ($Z_3$, $Z_2$ or at least a $Z_1$).
Now from Table 3 it is seen that one $Z_3$ along with one X-shift can accomplish $Z_3$ + X => (15 + 0.625)% = 15.625 % of mixing completion. Hence, remaining mixing to be accomplished = (100–15.625) % = 84.375 %.

Again from Table 3, it is evident; the fastest possible remaining mixing can be completed if the remaining mixing pattern consists of $Z_3$ and X-shift only alternately one followed by the other as shown in Fig 9. If such a shift-pattern can be set for the remaining mixing on the chip (without considering the routing congestion among different mixing paths), then total mixing completion in 1st cycle is computed by equation no. (3). In Fig. 9, $Z_3$ and X-shifts are respectively represented by Blue and Orange color in 1st cycle and by Black and Green arrow for the 2nd mixing cycle.

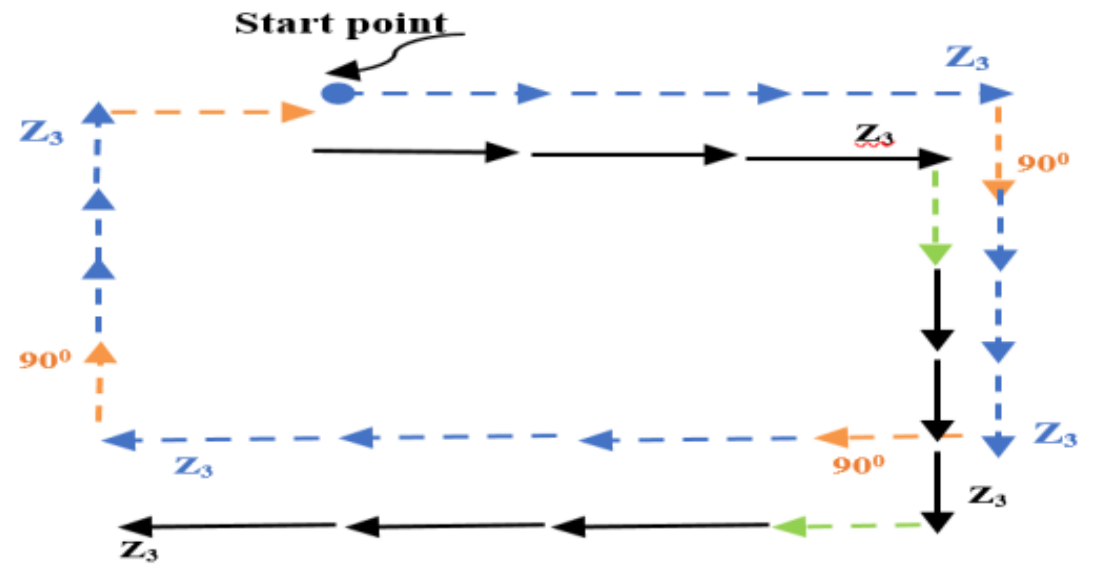

**Fig. 9**. Minimum steps required for mixing completion by MLS-CP

The mixing completion for 1st cycle is as follows:

$$Z_3 + X + Z_3 + X + Z_3 + X + Z_3 + X \Rightarrow 4\,(Z_3 + X) \quad (3)$$

$$\Rightarrow 4(15 + 0.625)$$

$$\Rightarrow 62.5\% \text{ of mixing}$$

The time required for 62.5% mixing is $4(3t) + 4(t) = 16t$. Where $t = 1$ time-step. Now to accomplish remaining mixing we need further movement as shown in Black and Green color. (2nd cycle)

$$3\,(Z_3\text{-shift}) + 2\,(X\text{-shift}) \Rightarrow 3\,(3t) + 2\,(t) = 11t \quad (4)$$

So total movement (time-steps) required is $= 16t + 11t = 27t$.

Now applying proving method of contradiction:

If we had chosen $Z_2$-shift or $Z_1$-shift in our shift-pattern i.e. from Fig. 9 if we delete one $Z_3$-shift (15% mixing completion) and in place of it we need to find three $Z_2$-shift to accomplish the same amount (15%) of mixing. But three consecutive $Z_2$-shift is not permitted and thus we need to put two X-shift in between them. Thus, total steps would be needed as follows:

$$Z_2 + X + Z_2 + X + Z_2 = 3\,(Z_2) + 2\,(X)$$

$$= 16.25\,\% \Rightarrow 3(2t) + 2(t) = 8t$$

Hence to accomplish approximately same mixing amount (16.25%) a single $Z_3$-shift can achieve in 3t we needed 8t (where t = one-time-step) and if we would have used $Z_3$-shift we need 5t for same amount 16.25%) of mixing completion.

Similarly, to accomplish the same amount of mixing using $Z_1$ in place of $Z_2$ we need

$$Z_2 = Z_1 + X + Z_1 + X$$

$$\Rightarrow 2\,(Z_1 + X) \Rightarrow 2(t+t) \Rightarrow 4t$$

So one $Z_2$ (which requires 2t) is equivalent to $2Z_1 + 2X$ which requires 4t time.

Hence using only $Z_3$ and mandatory X–shifts can produce the minimum t (time-steps) to finish the entire mixing and which is 27t as derived by Eqn. (3) and Eqn. (4) and thus can be concluded that below 27t it is not possible to accomplish total mixing (100% mixing completion) for a 16 Hz chip.

### B. *Synthesis Process Using Directional-Shift*

We perform the assay-synthesis using above-mentioned shift-movements ($Z_3$, $Z_2$, $Z_1$, X, Y) rather assigning a dedicated module. The proposed new methodology of Module-Less-Synthesis for Cyber physical DMFB (MLS-CP) uses the above-derived shifts and finds out paths (patterns) for each mixing operation until 100% mixing (dilution) is achieved. Thus, eventually, we have converted the synthesis problem into a routing problem and solved with our proposed directional-shift (movements) of droplets considering all the routing constraints on a DMFB platform.

The MLS-CP method starts with the initialization of the coordinates for all mixing operations at first level of sequence graph (G). Usually, the top and bottom corner positions of the chip are chosen to start the first stage of mixing (for $\leq 4$ number of mixing). Next, it searches for all possible shifts from the initial coordinate position. It first checks for the availability of linear-shifts ($Z_1$, $Z_2$, $Z_3$). If no Z-shift is available, $90^0$(X-shifts) are searched. The parallel movement of mixing-droplets on the cyberphysical DMFB is always done by checking the routing constraints at each time-step and the percentage of mixing completion is updated accordingly in the control software. Thus to synthesize the bio-protocols we have used the proposed shift patterns and we set the precedence order of these shift-patterns. The proposed MLS-CP algorithm is given in Fig. 10.

According to MLS_CP, if a $Z_1$-shift is available, then the droplet tends to achieve straight run movements ($Z_2$) or a $Z_3$-shift in a greedy manner or until a collision occurs with other droplets. As depicted in Lemma 3; achieving more number of $Z_3$-shifts (if available) results in faster completion of the mixing operation.

***Lemma 3:*** **In MLS_CP Precedence Order set for various Shift-Movement is: $Z_3 > Z_2 > Z_1 > X > \text{STALL} > Y$**

***Proof.*** From Fig. 9, it is now evident that $Z_3$-shift is having the highest mixing completion percentage (15%). If one $Z_3$ is deleted it can be replaced with one $Z_2$ and one X move such that

$$Z_3 \equiv Z_2 + X \quad (5)$$

By fixing the same number of time-steps (here 3t) the mixer droplet can have either $Z_3$ or $[1(Z_2) + 1(X)]$ movements (shifts). Now one $Z_3$–shift accomplishes 15% mixing as given in Table 3. But one $[(Z_2) + (X)]$ combining accomplishes $[5+0.625] = 5.625\%$ mixing. Hence, $Z_3$-Shift should be given higher precedence order than $Z_2$ i.e.

$$Z_3 > Z_2 \quad (6)$$

Similarly fixing the time-step (t) on the chip one $Z_3$ can be replaced with one $Z_1$ –move and two X-move such that

$$Z_3 \equiv 1(Z_1) + 2(X) \quad (7)$$

Nevertheless, $Z_3$ completes 15% whereas $[Z_1 + 2(X)]$-shifts combining accomplishes $[1.875 + 2(0.625)] = 3.125\%$ of mixing. Hence,

$$Z_3 > Z_1 \quad (8)$$

Similarly $Z_2 \equiv 1(Z_1) + 1(X)$ but $Z_2$ completes 5% mixing whereas $[1(Z_1) + 1(X)]$ combining finishes $[1.875 + 0.626] = 2.5$ % of mixing. Hence $\quad Z_2 > Z_1$ (9)

So from Equation No. (6),(8) and (9) we can conclude that Precedence of shifts will be as follows

$$Z_3 > Z_2 > Z_1 \tag{10}$$

Similarly fixing the time for 3t duration we can have either $Z_3$ or 3(X) s.t. $Z_3 \equiv 3(X)$ and using induction it can be proved

$$Z_3 > X \tag{11}$$

Similarly, by principle of induction $Z_2 \equiv 2(X)$ and $Z_1 \equiv 1(X)$ in terms of time-steps required to perform the shifts. Thus, we can conclude the precedence order of our proposed shifts (movements) with respect to mixing accomplishments are as

$$Z_3\text{-}shift > Z_2\text{-}shift > Z_1\text{-}shift > X\text{-}shift > Y\text{-}shift \tag{12}$$

The precedence of STALL operation is set in between X and Y-Shift. By experimental analysis on various chip sizes and dimension, in many a case we came across typical deadlock situation where no further shift movement of the droplets are possible. To avoid such scenario we introduce the "STALL"-movement. It is actually stalling of the mixer-droplet for a single time-step (t=1) to pass the other droplets. Theoretically, 'STALL'-movement can accomplish 0% mixing completion. Hence, the priority of 'STALL' is set higher than Y-Shift, which have negative mixing accomplishment (-2.5%). The Y-shifts are only used whenever 'STALL' operation is performed for 3 consecutive time-steps for a mixer droplet and irrespective of that deadlock situation holds.

Hence, priority wise the following relation holds true for MLS-CP method and the droplet will move in accordance.

**$Z_3$–shift>$Z_2$-shift>$Z_1$-shift>X-shift>STALL>Y-shift**

The pseudocode for Split_and_Merge and MLS_CP_Synthesis algorithms are given in Fig. 10(a) and in Fig. 10(b) respectively.

If an X-shift is to be obtained, then Left and Right distance to the boundary from the current coordinate position is checked. If routing constraints on both sides are satisfied, then the droplet is moved in the direction of maximum length available. However, if both the Left and Right distances are equal (Line No. 20-22, Fig.10b), the droplet is moved towards the droplet, which will co-parent its child in the next level of sequencing graph.

If routing constraints are not satisfied in either direction, then the droplet is STALL-ed in its current position. We assume 0% mixing completion takes place for stalling operation. If a droplet is STALLED for 3 time-steps consecutively (line no. 38-43 of Fig.10.b), it is assumed to be in deadlock and then only it moves in an $180^0$ direction (Y-shift) to break the deadlock. In general, Y-shifts are avoided because of its negative mixing completion (-2.5%).

In the Split_and_Merge Algorithm (Fig. 10.a), *SPLIT_SET's* are formed for all completed mixing operations. When fully mixed droplets are at corner electrode, there will be 3 coordinates at max in the *SPLIT_SET*. Else, depending on the

**Module-Less-Synthesis Algorithm for Cyberphysical DMFB (MLS-CP Algorithm)**

***Input:***
1. ***Sequence Graph (G)***
2. ***Chip Architecture (X, Y)***
3. ***THRESHOLD16***

**Algorithm 1: Split_and_Merge ($M_j$)**

1: Check Co-ordinates $\forall M_j$ ;
2: Form Probable SPILIT_SET μ considering Routing Constraints;
3: **for** $M_j \in L_{k+1}$ **do**
4: $\quad \delta_{min} \leftarrow$ **Min** (Manhattan Distance among all $(x_i,y_i) \in \mu_L$ and $(x_i,y_i) \in \mu_R$);
$\quad$ Where, μL= SPLIT_SET of Left Parent node; μR= SPLIT_SET of Left Parent node
5: $\quad$ ***SET_TARGET_COORDINATE*** $\leftarrow \dfrac{\delta_{min}}{2}$
6: $\quad$ Perform Routing of Droplets to the *TARGET_ COORDINATE*
*7:* ***end for***
8: Start module less Synthesis Algorithm for next level mixing operations $M_{j+1} \in G$
*9:* ***end procedure***

**Fig. 10(a):** Pseudocode for Split_and_Merge

droplets coordinate position (boundary electrode cell /intermediate cell) there will be 2 or 4 coordinates at max in the *Split_Set*. Manhattan distances are calculated from all the coordinates (x,y) of the *Spit_Sets* of the co-parents and the least distance is found out. An electrode at distance *Min {(Manhattan Distance*) / 2} is chosen as target cell and the corresponding droplets are routed to that target electrode (line no. 5-8, Fig. 10.a). The Module-less mixings start again for next level of synthesis according to the sequence graph (G) and the process continues till the entire assay synthesis ends.

## C. *Congestion Avoidance in MLS Method*

In MLS-CP method, entire mixing operation has been converted into routing steps of variously proposed shift-patterns (*$Z_3$, $Z_2$, $Z_1$, X, Y and STALL*) where the mixer-droplets traverse on the chip according to the precedence order set by Lemma 3. One shortfall of this method is it will essentially cause higher congestion on the chip with time and thus finding various shifts become difficult as time progresses. We have found the parallel mixing operations incurred more number of STALL operations at the later stage of synthesis in our proposed MLS method and the available route decreases as time t increases.

To avoid such congestion and to reduce the overhead washing operation we have introduced a Modified model of Module-Less-Synthesis (MMLS).

In this new MMLS approach, we have introduced the concept of chain formation by various shift-movements as shown in Fig. 11. We assume all the mixer droplets $M_1$, $M_2$ ....$M_5$ starts simultaneously on the 8x8 chip and out of these 5-mixer droplets $M_1$, $M_2$ and $M_3$ are able to form a chain on the chip shown by Red, Blue and Green colour. Droplet $M_4$ and $M_5$ were not able to form the chain as there was not enough space on the chip to take two consecutive X-Shifts on 4th and 5th time-step and hence they follow earlier MLS shift-movements.

To form the chain the droplet need to take two consecutive X-Shift ($90^0$-shift) on 4th and 5th time-step even though it might have the scope of taking a better shift-movement ($Z_3$ / $Z_2$ / $Z_1$) which would have accomplished more mixing percentage at the end of 5th-time step. Taking 2 consecutive X-shifts provide the opportunity the get one more $Z_3$–shift on 6th, 7th, 8th time steps which ultimately accomplish much faster mixing. Therefore, forming a chain might seem to be disadvantageous but over the long run, it becomes beneficial compared to traditional MLS process, as the chip-space used for that particular mixing operation is much lesser (5x2 = 10 cells only) and faster mixing is accomplished. As soon as we are able to form the chain, all cells consisting the chain will be blocked (5x2 = 10 cells only) for the entire duration of the mixing (presumed to be 32t where t = 1 time-step with a chip operating frequency f = 16 Hz and t = 1/f). Thus, for up to 32 time-steps, no other malicious droplet can intrude the chain and there is no scope of cross contamination with other droplet. Also shift movement complexity of the mixer droplet reduced within the chain as the sequence of shift-movements get fixed within a chain, which is as follows:

$$(Z_3) \rightarrow (X) \rightarrow (X) \rightarrow (Z_3) \rightarrow (X) \rightarrow (X)$$

The above shift-pattern is the same as deduced in Lemma 1 and the chains (Fig. 11) function similar to a 2x5 module.

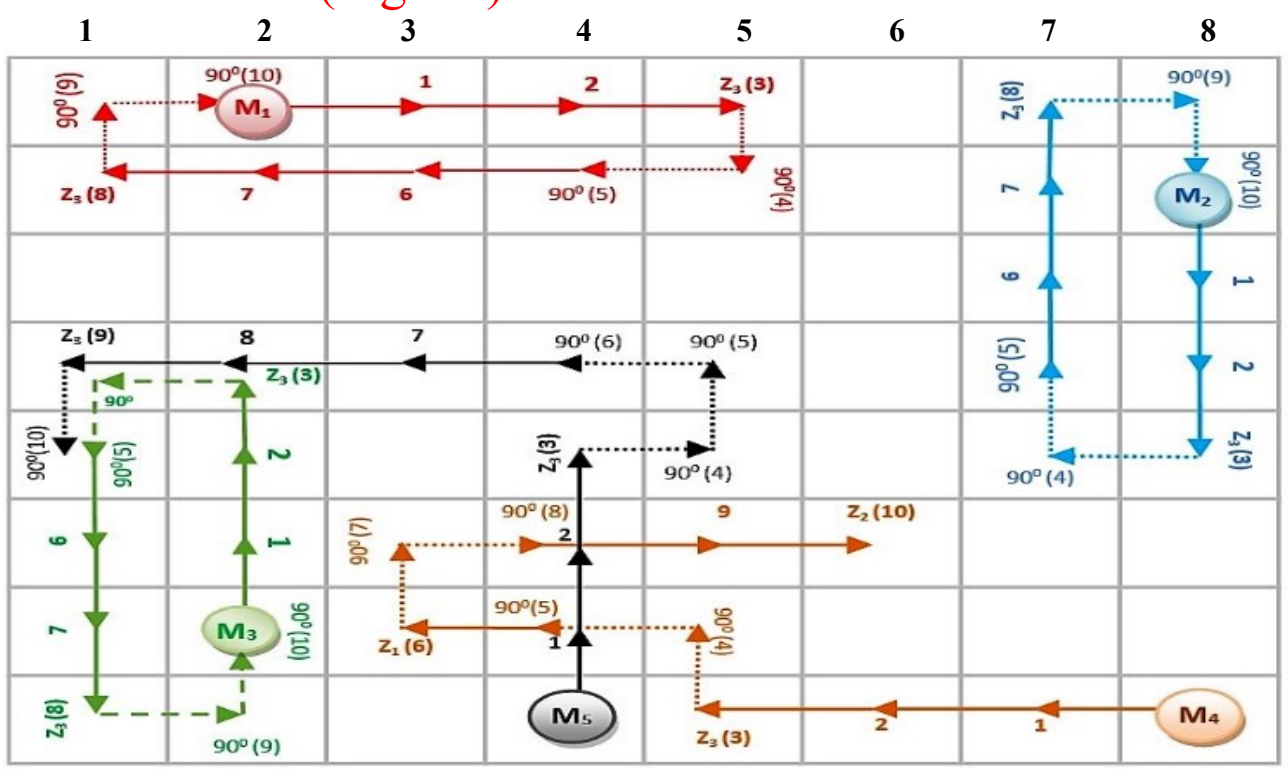

**Fig. 11a:** Modified Module-Less Mixing Paths for $M_1$ to $M_5$ till 10th time step
***1-time step** = 0.0625 Sec by considering CP DMFB working frequency f = 16Hz.

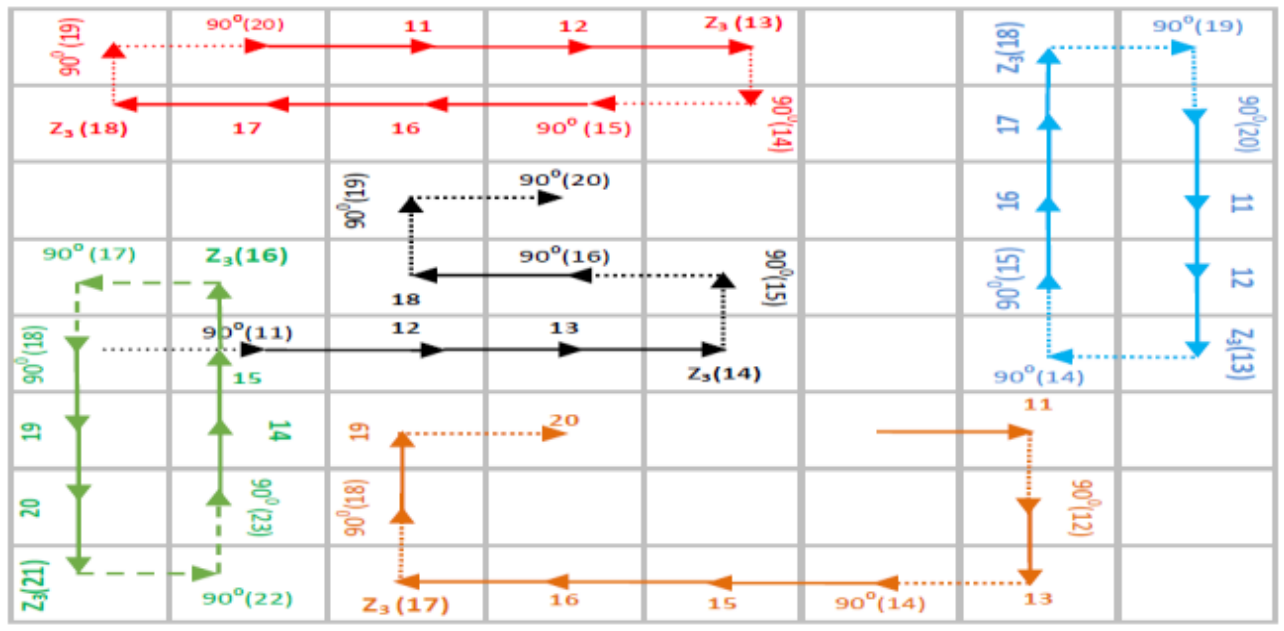

Fig. 11.b Modified Module-Less Mixing Paths for $M_1$ to $M_5$ till 20th time step

**Algorithm 2: MLS_Synthesis Algorithm**

```
Procedure MLS_CP
  Check If (Td = = 16)
    GOTO Error_Checking_Module ( );
  Else
for all levels L ∈ G do
   Identify all mixing operations Mj ∈ Lk;
   initialize co-ordinates ∀ Mj;
   Z_COUNTER ← 0; X_COUNTER ← 0; STALL_
            COUNTER ← 0; STEP_COUNT ← 0;
   η ←Select from SHIFT_MOVEMENT_SET ρ = (Z,X, STALL,
   Y) maintaining Routing Constraints;
switch η do
case Z (0°) :
  UPDATE STEP_COUNT:
   if Z_COUNTER == 3 then
      Z_COUNTER ← 0;
      GOTO Case (X)
                  end If
if Z_COUNTER ≠3 then
   X_SHIFT_COUNTER ++;
   Perform X-Shift;
  UPDATE COORDINATES (Mj) and PERCENTAGE _COMPLETION (Mj);
                end if
case ( X) :
               Calculate LenR, LenL;
if Routing Constraint Satisfied in both Direction then
   if (LenR = = LenL) then
      Choose X-shift according to Mj ∈ Lk+1;
   end if
  If LenR ≠ LenL then
   Move in the direction of MAX (LenR, LenL)
  end if
  if Routing Constraints on LEFT side satisfied
     Move in the direction ← LEFT ;
   end if
  if Routing Constraints on RIGHT side satisfied
    Move in the direction ← RIGHT ;
   end if
end if
UPDATE COORDINATES (Mj) and PERCENTAGE _COMPLETION (Mj);
 case (STALL)
   if STALL_COUNTER == 3 then
      STALL_COUNTER = 0 and Perform Y-Shift;
   end if
  else
  STALL_COUNTER ++;
End for
  Call Split_And_Merge Algo. ∀ Mj ∈ Lk
End procedure
```

**Fig. 10(b):** Pseudocode for MLS-CP Synthesis

As a result, we can omit the washing process within the chain (which are of size 2x5) as the mixer droplet is homogeneous in nature and there is no need to wash until the entire mixing gets finished within a chain. Only for the Non-Chain mixing operations ($M_4$ and $M_5$ in Fig. 11), we need to consider the washing overhead at each time-steps. The MLS_washing algorithm is shown in Fig. 10(c).

Therefore, the MMLS process is somewhat a combined approach of module-based mixing and MLS mixing strategies, incorporating advantages from both of them. From Fig. 11, it is seen for $M_1$, $M_2$ and $M_3$ mixing- operations, chain can be formed which is somewhat similar to Module based mixing but without the requirement of padding cells all around the chain. Thus, a lot of chip space can be saved as most of the mixing operation usually starts from the boundary cells closer to the chip reservoirs. $M_4$ and $M_5$ droplets are unable to form a chain and thus they follow traditional MLS shift-patterns as shown by Orange and Black colour respectively.

### D. *Satisfiability Solver Model for Deadlock Avoidance in MLS Method*

It is seen that up to 5 number of parallel mixing on an 8x8 chip, we are able to find the solution but as parallel mixing operation increases finding paths on the chip becomes difficult. To avoid such problem for a higher number of parallel mixing operations (≥ 6) we form a satisfiability problem of such overlapped droplet movement and run those using SMT algorithms. SMT can handle instances composed of hundreds of thousands of variables and constraints in a reasonable time [24]. As each instant of time 't' which droplet should move in which direction out of four possibilities on a DMF chip is a decision problem and iteration of such decision problems over many instances of time actually formed the optimization problem for finding minimal route for all the conflicting droplets shown in Fig. 12. The problem can be formulated using four functions in disjunctive normal form as below

$$\Phi = F1 \vee F2 \vee F3 \vee F4 \quad (13)$$

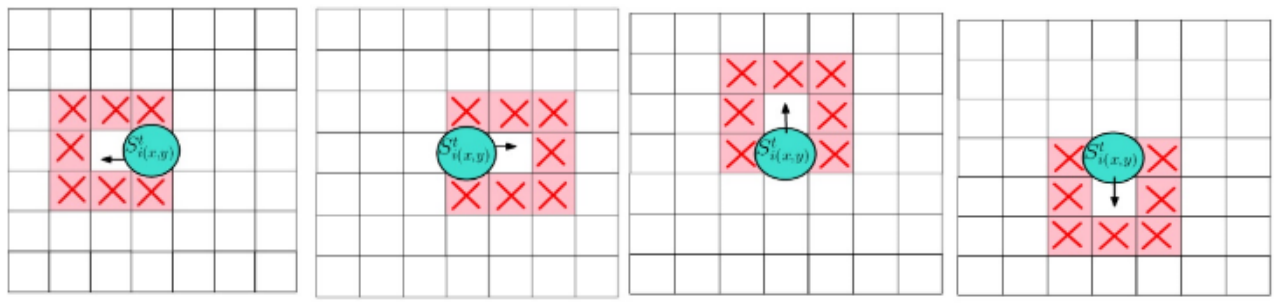

a) LEFT move b) RIGHT move c) UP movement d) DOWN movement

**Fig. 12** Permissible droplet movements in terms of satisfiability problem

Where $F_1$, $F_2$, $F_3$, $F_4$ represents left, right, up and down movement of the droplet, respectively, and each of them can be represented as a conjunctive normal form s.t.

$$\exists F_1 S^t_{i,x,y} \xrightarrow{LEFT} \neg S^t_{i,(x-2,y)} \wedge \neg S^t_{i,(x-2,y+1)} \wedge \neg S^t_{i,(x-2,y-1)} \wedge \neg S^t_{i,(x-1,y+1)} \wedge \neg S^t_{i,(x-1,y-1)} \wedge \neg S^t_{i,(x,y+1)} \wedge \neg S^t_{i,(x,y-1)} \quad (14)$$

$$\exists F_2 S^t_{i,x,y} \xrightarrow{RIGHT} \neg S^t_{i,(x+2,y)} \wedge \neg S^t_{i,(x+2,y+1)} \wedge \neg S^t_{i,(x+2,y-1)} \wedge \neg S^t_{i,(x+1,y+1)} \wedge \neg S^t_{i,(x+1,y-1)} \wedge \neg S^t_{i,(x,y+1)} \wedge \neg S^t_{i,(x,y-1)} \quad (15)$$

$$\exists F_2 S^t_{i,x,y} \xrightarrow{UP} \neg S^t_{i,(x,y+2)} \wedge \neg S^t_{i,(x-1,y+2)} \wedge \neg S^t_{i,(x+1,y+2)} \wedge \neg S^t_{i,(x-1,y-1)} \wedge \neg S^t_{i,(x+1,y+1)} \wedge \neg S^t_{i,(x-1,y)} \wedge \neg S^t_{i,(x+1,y)} \quad (16)$$

$$\exists F_2 S^t_{i,x,y} \xrightarrow{DOWN} \neg S^t_{i,(x,y-2)} \wedge \neg S^t_{i,(x-1,y-2)} \wedge \neg S^t_{i,(x+1,y+2)} \wedge \neg S^t_{i,(x-1,y-1)} \wedge \neg S^t_{i,(x+1,y+1)} \wedge \neg S^t_{i,(x-1,y)} \wedge \neg S^t_{i,(x+1,y)} \quad (17)$$

Thus using SMT SAT Solver, we are able to get generalized paths for the mixer droplets and based on those available paths we have applied our proposed Shift-patterns to accomplish the mixing operations. For few instances, we have incurred deadlock situation but applying "STALL" operation, which is considered as one of the Shift-movement we are able to avoid all the deadlock faced in complex benchmarks (Benchmarks assumed with ≥ 6 parallel mixing operation in a single layer) also. The obtained results after applying SMT Solver on MMLS are given in Section VII.

## V. Error Recovery in MLS

We have introduced an efficient error detection strategy to ease the recovery process and applied it successfully on our proposed Module Less Synthesis (MLS) /MMLS technique. It guarantees 100% mixing as well as synthesis completion on cyberphysical DMFB platform with much lesser time required for recovery in case of faults arises on the chip. The detailed error recovery strategy is discussed below.

### A. *Error Types and Chip Reliability*

We consider the types of errors and principles of recovery for such errors on a cyberphysical DMFB. For a given synthesis performed by MLS, the primary fluidic operations on cyberphysical DMFB and their respective strategies are given in Table 4. In this work, we have pre-assumed CCD camera-based sensing integrated with our cyberphysical chip for its on-spot faster detection time and flexibilities compared to optical detector units [20]. After detecting an error, two different strategies may be adopted for re-execution of the erroneous operation.

I. The operation with an error is re-executed and it is assumed that the same MLS/MMLS mixing-patterns (paths) on the chip will be followed during the re-execution of the operation.

II. The entire chip can be divided into several clusters. The erroneous electrodes will not be used at the time of re-execution. The control software keeps track of all the on-chip resources occupied by a particular operation. Thus, depending on the erroneous droplet, it is easy to backtrack to the region where an error occurs. We consider all electrodes in this region as the possible locations for defects, avoid the entire region next time onwards, and try to adapt next best combinations of available MLS shifts (Paths) which were not used in earlier run.

**Table 4**: Recovery strategies for various DMFB Operations [20]

| Reversible Operations | Recovery Strategy | Non-reversible operations | Recovery Strategy |
|---|---|---|---|
| Dispensing, Splitting | REPEAT Operation | Mixing, Dilution, detection via optical detector | Need droplet from Previous level / Backtracking |

### B. *Recovery Problem in MLS*

In our example problem of Module less Synthesis approach, operating- frequency of the chip is taken as 16 Hz. Hence 1 time-step (t) = (1/16) = 0.0625 sec. The recovery problem is formulated as to find the optimum time step ($t_d$) at which CCD based monitoring (checkpoint) should be placed so that the recovery overhead (in terms of the traversed path) will be minimum.

If we consider MLS /MMLS shift-movements (patterns), in case of Chain mixing the recovery is same as module based mixing i.e. after completion of entire mixing we can only check whether the mixing accomplishment successful or not. To determine mixing completion we need to wait until the end of mixing operation. However in case of Non-Chain mixing patterns we propose the new scope of improvement for error detection. Fig. 14.a, 14.b shows various directional shifts for PCR assay according to MLS pattern. Experimenting MLS on all the available DMFB benchmarks and on other synthesized hard test benches [24], the following assumptions are taken into consideration.

***<u>Assumptions</u>***

I. In MLS approach, the average time required to accomplish 100% mixing is 32t, i.e. the average path to be traversed by a mixer-droplet is fixed to be 32 to complete the mixing.

II. The entire path ($n_0$, $n_1$, $n_2$…..$n_{31}$) consists of all 32 distinct cells (Non-Chain mixing) on the chip traversed from $t_0$ time steps to $t_{31}$, i.e. no cell is repeated twice on the path (P) as shown in Fig. 11 and no STALL operation is required in these 32 steps.

III. The entire chip is assumed a 'fair-chip'. Hence, the probability of each cell on the entire path length (P =32) being defected to be equal to exactly same to the other cells belongs to the path i.e. probability of each cell being defected on the path is 1/P.

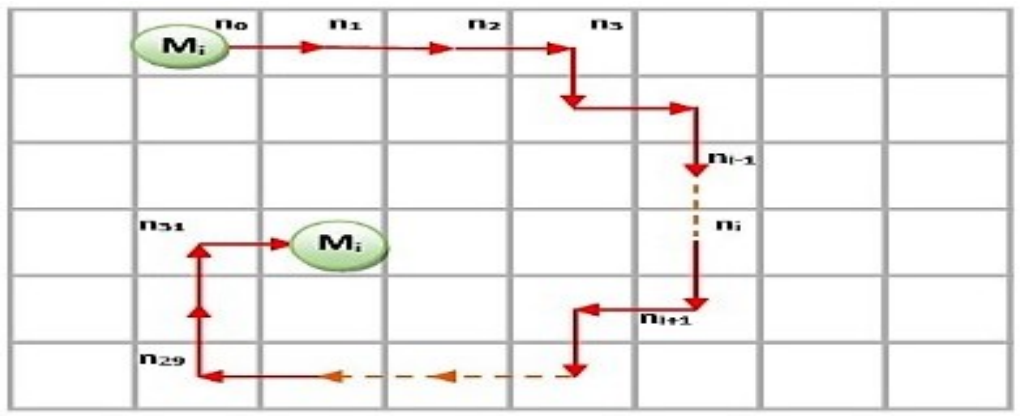

**Fig. 13.** Average path Length (P) of 32 cells

### C. *Error Detection Strategy for MLS Method*

Integrating physical-aware software with MLS method, which can read and analyze sensor data and dynamically adapt the given synthesis, our technique can be updated with modified sequence graph, scheduling operations, and droplet routing pathways in runtime. Unlike [20], MLS control software does not need to consider module placement and resource binding phase for each of the fluidic operations and thus it minimizes initial synthesis time as well as online re-synthesis time after monitoring of the assay, if needed.

Here we do not target to monitor the cyberphysical chip for every time-steps (t) of the entire synthesis duration, which is expensive and computationally intensive for the control software. Checking at each time-stamp (t) would add huge overhead time to the assay execution due to the dynamic re-synthesis of the operations as and when needed. Such an error recovery strategy ultimately slows down the entire synthesis. Hence, we have proposed to monitor the chip only at a single intermediate time instant ($t_d$) during the entire operational (mixing) time for once and after completion of the mixing operation the final output will be monitored at ($t_f$) for once only. Thus, incorporating only one intermediate checkpoint with the traditional error recovery strategy will minimize the re-synthesis overhead at runtime and makes entire error recovery much faster for MLS approach than conventional module based synthesis. If we consider at time-step $t_d$ the CCD camera gets activated and image snapshot is taken to monitor the droplet then we can easily rollback the operation from $t_d$ instead of waiting till the finish time ($t_f$ ) of the operation and then rolling back. For example if we monitor as early as at $8^{th}$ time step i.e $t_d = 8$ and $1 \geq t_d \geq 32$ and if we are able to detect error at $t_d$, we shall not continue the operation anymore till the end ($t_f$) and roll back from $t_d = 8^{th}$ time-step itself. If we assume the next run to be fault free and the said operation finishes successfully at $t_f$ ($t_f = 32$) in next run. The total time required to finish the operation would be:

$$T_d + T_f = 8 + 32 = 40t$$

Now according to ***Assumption No. II*** of V.(B) of previous subsection, the number of faults which can be detected at intermediate checkpoint ($t_d$) would be 8 and the remaining 24 faults ( for cell number $n_9$ to cell number $n_{31}$) cannot be detected at intermediate checkpoint ($t_d$) here and those may only be detected at final checkpoint ($t_f$) i.e. after finishing of the operation. Because of equally likely probability (Assumption No. III of V.(B)) of having an individual cell being defected on the entire path length, the total time ($T_R$) required would be as follows for large number of trials.

$$T_R = \textit{(Nos. of faults detected at } T_d\textit{)} \; \mathbf{x} \; \textit{(Recovery time from } T_d\textit{)} + \textit{(Nos. of faults detected at } T_f\textit{)} \; \mathbf{x} \; \textit{(Recovery time from } T_f\textit{)} \quad ...(18)$$

With above example if $T_d = 8$ and $T_f = 32$ ;

Total time-steps ($T_R$) required recovering from the errors would be as follows:

$$T_R = 8 \text{ x } (8 + 32) + 24 \text{ x } (32+32)$$
$$= 8 \text{ x } 40 + 24 \text{ x } 64 = 1856t \quad (t = 1 \text{ time-step})$$

Similarly, if we fix the intermediate detection at $T_d = 10$ and $T_f = 32$

$$T_R = 10 \text{ x } (10 + 32) + 22 \text{ x } (32+32)$$
$$= 1828t$$

Putting different numerical values in $T_d$ we get the parabolic curve as shown in Fig. 14. It is seen that if the early intermediate detection checkpoint is fixed at $T_d = 16$ time-step i.e. exactly at the halfway path, the overall recovery time can be minimized. So we fix the detection twice, first at $t_d = 16t$ after the mixing operation gets started and finally at completion ($T_f = 32t$) i.e. after finishing of the entire operation.

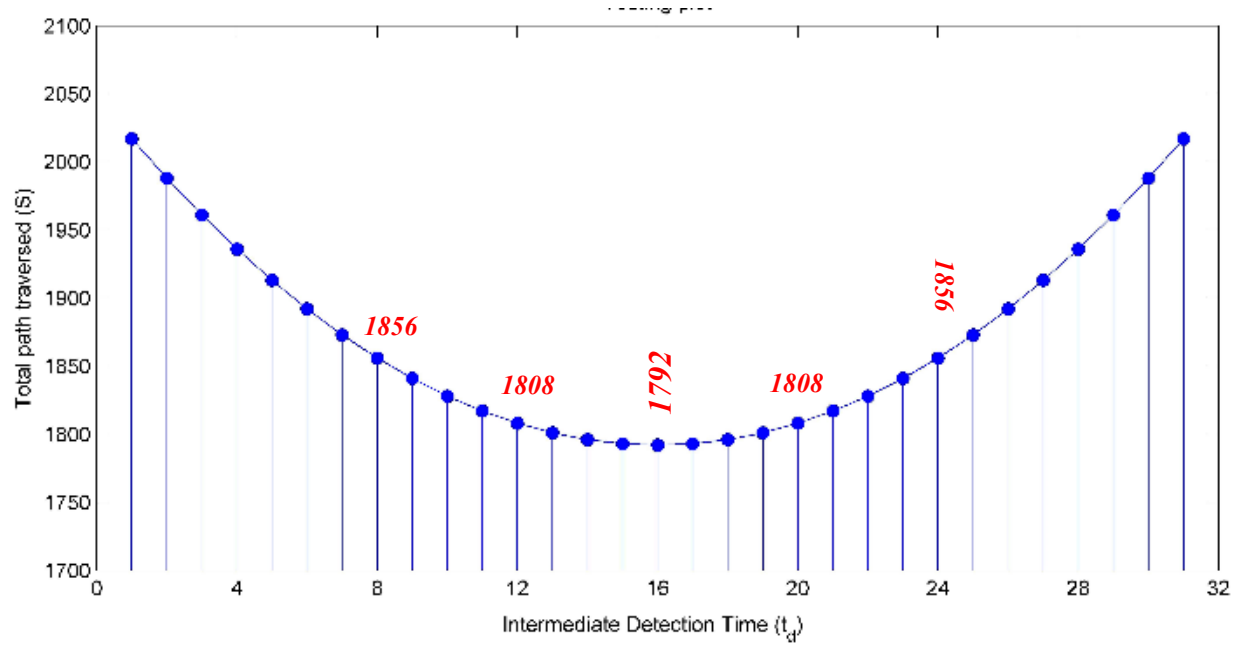

**Fig. 14.** Early Detection time ($T_d$) vs. Recovery time / Total Path Traversed (S)

***Lemma 4:*** **For MLS, the error recovery time will be minimized if the intermediate error detection checkpoint is fixed at n= N/2 time-steps, where N is the total path length.**

To generalize the above concept we assume:
N = Total Path length;
n = intermediate single checkpoint
S = total path travelled to complete a mixing operation including rollback

Now we simulate the same mixing operation for a very large number of trials. According to the theory of classical probability, 'n' number of faults may happen at first 'n' cells on the path and can be detected at $T_d$ (Fig. 13). The remaining (N-n) number of faults will happen on rest of the path length and can be detected at final checkpoint ($T_f$) only.
For first 'n' number of faults on the MLS-path the distance traversed by the droplet is:

$$S_1 = (n + N)\, n \qquad (13)$$

For remaining (N-n) faults the total distance to be traversed by the droplet is:

$$S_2 = (N + N)\,(N-n) \qquad (14)$$

So the total path (S) calculated from Eqn. No.(13) and (14) :

$$S = S_1 + S_2$$
$$\Rightarrow (n + N)\, n + (N+N)(N-n)$$
$$\Rightarrow n^2 + Nn + + 2N\,(N-n)$$
$$\Rightarrow n^2 + Nn + 2N^2 - 2Nn$$
$$\Rightarrow n^2 - Nn + 2N^2 \qquad (15)$$

Eqn. no (15) is a symmetric function and we need to find the value of 'n' for which the total recovery path as well as operation completion path (S) should be minimum. Thus finding derivative from equation No. (15)

$$(ds / dn) = 2n - N$$

Putting, $(ds / dn) = 0$; we find $n = N/2$; and $d^2s / dn^2 = 2$; which is a positive integer and sufficient condition for existing minima at n = N/2. Hence, the above claim of n= N/2 is proved i.e. we fix the intermediate detection (checkpoint) exactly at halfway path of the proposed MLS-shifts.

If a certain mixing operation fails to achieve required threshold values (threshold set according to sequence graph of the biochemical assay and nature of biochemistry) at time step n=16, then that particular mixing operation is roll backed without interrupting others. Thus, MLS_method does not check for a threshold value of mixing completion at higher time steps and does not wait for rollback till the entire mixing is complete, unlike module-based synthesis. Here errors are detected much faster and rollback operation can be started at an earlier stage (at t=16). The unused droplets at the previous level are stored and can be used for faster rollback till the next level of mixing is complete. These stored droplets are discarded to the waste reservoir while the next level of mixing achieves 100% completion. This process continues until the entire bioassay synthesis is complete.

## VI. Illustrative Example

### *A. Mixing Stages for PCR by MLS method*

The mixing stages of PCR shown in Fig. 3(a), consists of seven mixing operations denoted by $M_1$ to $M_7$. The path obtained by MLS method for simultaneous mixing operations ($M_1$, $M_2$, $M_3$ and $M_4$, which are shown by Blue, Yellow, Green and Red color respectively in Fig. 14 for an 8x8 chip size. Fig. 14(a) and 14(b) shows the MLS shift-movements (path) obtained till 17th and 30th time-steps respectively. The corresponding mixing completion percentages are computed and shown below. The associated coordinate positions written in subscript are the final placement of the droplet on the chip and RHS value signifies respective mixing completion percentage till 17t where t=one time-step.

- $M_1 = Z_3 + X + Z_1 + X + Z_2 + X + Z_3 + X + Z_3 + X$
  $M_1 = 3 * Z_3 + Z_2 + Z_1 + 5 * X$
  $M_1 = 3\,(15\%) + (5\%) + (1.875\%) + 5\,(0.625\%)$
  $M_{1\,(6,\,5)}$ = **55 %**
- $M_2 = Z_3 + X + Z_1 + X + Z_2 + X + Z_3 + X + Z_3 + X$
  $M_{2\,(4,\,6)} = 3 * Z_3 + Z_2 + Z_1 + 5 * X$ = **55%**
- $M_{3\,(3,4)} = Z_3 + X + Z_1 + X + Z_2 + X + Z_3 + X + Z_3 + X$ = **55%**
- $M_{4\,(5,\,1)} = Z_3 + X + Z_3 + X + Z_2 + X + Z_2 + X + Z_3$ = **57.5%**

The above computation continues until full mixing is accomplished for $M_1$, $M_2$, $M_3$ and $M_4$. The corresponding mixing completion (100%) and their respective coordinate positions as shown in Fig. 14. (b) Which are $\mathbf{M_{1\,(2,5)}}$, $\mathbf{M_{2\,(4,2)}}$, $\mathbf{M_{3\,(8,4)}}$, and $\mathbf{M_{4\,(5,6)}}$. It is observed that $M_1$ and $M_2$ takes 29 time steps(t) to accomplish full mixing (100% completion) with their corresponding (X, Y) coordinate positions are at (2,5) and (4,2), whereas $M_3$ and $M_4$ take 30t to accomplish full mixing. In the 30th time-step, $M_1$, $M_2$ and in 31st time-step $M_3$, $M_4$ split according to the *Split_and_Merge* algorithm and shown below in Fig. 15.

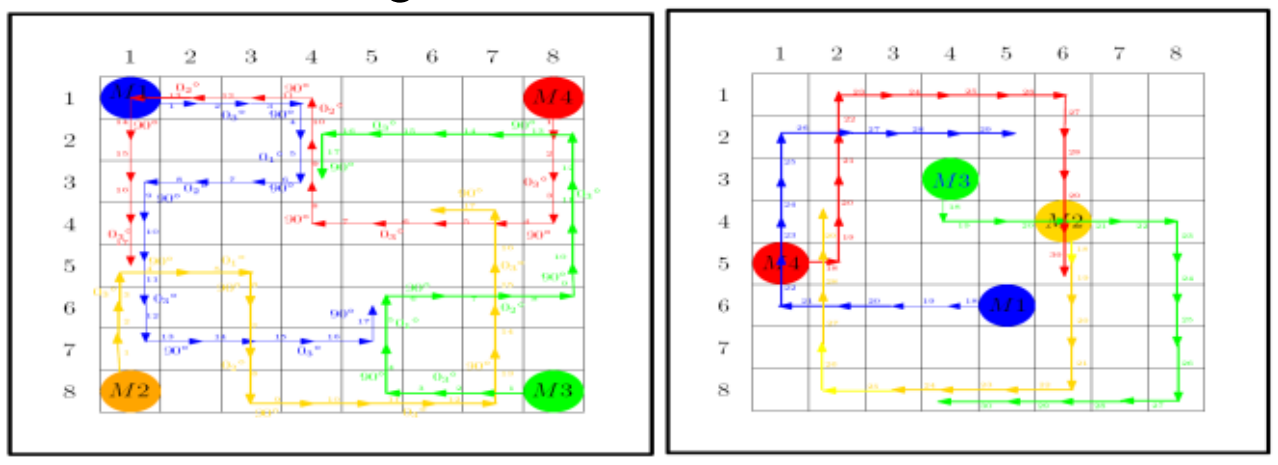

**(a)** till 17 steps (1.0625 sec) **(b)** till 30 steps (1.875sec)

**Fig. 14:** Module-Less Mixing Paths for PCR Bioassay

After 32t, next level of mixing operations $M_5$ and $M_6$ can be started as per the sequence graph of PCR and completed at 59t, then again the Split_and_Merge of 2nd level of PCR will be done. Similarly, the merging of $M_5$ and $M_6$ to form $M_7$ is done (ended at 65t) followed by the final mixing operation $M_7$ which needs further 27 steps. Thus, the PCR is completed at 92nd time-steps as shown in fig. 16, and total time required to finish the PCR assay is (92 x 0.0625) = 5.75sec which is approximately 40% improvement compared to earlier module based methods in our example problem.

Similarly, we have applied MLS methos on other available benchmark assays [24] including some hard test benches [24] and the obtained result is quite significant.

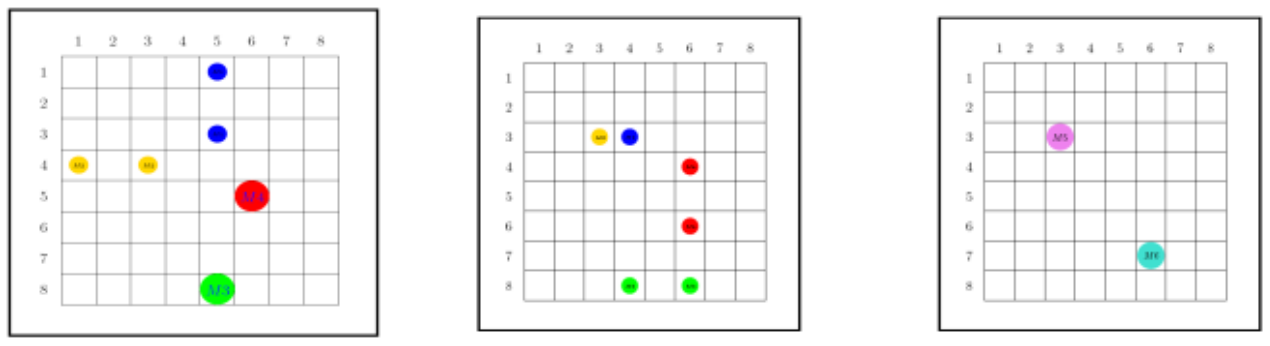

**(a)** Splitting of *M*1 and *M*2 at 30th time-step (t) **(b)** Split and Merge after 31st time-step **(c)** Split and Merge after 32 time-steps

**Fig. 15:** Split and Merge steps after completion of Level 1 of PCR

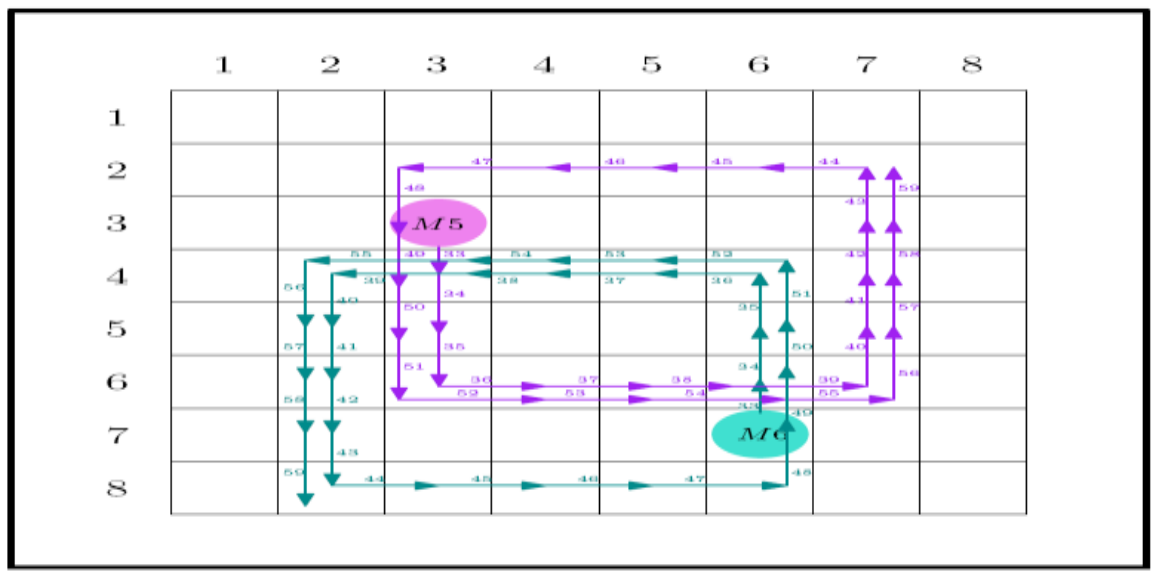

**Fig. 16:** Module-Less Mixing Paths for $M_5$ and $M_6$

*** t = 0.0625 sec by considering working frequency be 16Hz.**

## VII. Experimental Results

In order to evaluate the proposed MLS algorithm in Cyberphysical DMFB, two types of simulations have been performed. Primary focus has been given on determining the completion time improvements by MLS compared to the existing module-based synthesis techniques [26] using PCR and In_vitro Diagnostics (IVD) tests. Secondarily MLS method is applied on different available benchmarks to verify successful synthesis completion of the test cases in reasonable time for cyberphysical DMFB's. The simulation program is implemented in C++ language and SMT-Z3 solving engine is used as satisfiability solver with the open source toolkit metaSMT [27] on a 2.4 GHz Intel core i3 (M370) machine with 8 GB of memory. The findings are quite encouraging and discussed below.

### A. Results on IVD and PCR Benchmarks

The MLS synthesis results on IVD of physiological fluids and PCR benchmarks are compared with other module-based synthesis based on ILP and Tabu-Search [12]. For different chip dimensions and sizes (56 cells to 72 cells), bioassay synthesis time is promisingly improved around 35% by MLS approach as shown in Table 4.

### B. Chip Size Vs. Completion time in MLS

The scalability of the proposed method has also been tested by varying the chip size from 4*x*4 to 8*x*8 (16 to 64 cells) for PCR bioassay and it is observed that the completion time varies with chip size. Depending on chip dimension, if more $Z_3$ shifts are available and droplets get lesser or no STALL operations, faster mixing is achieved. Table 5 shows mixing completion times for different chip sizes. We run the 4 parallel mixing operations of PCR (stage I) on different chip sizes starting from 4x4 to 16x16. For 4 number of parallel mixing operations completion-time decreases till the chip size reaches 8x8 but after that increment of chip size effects very marginal decrement of completion. So we can conclude that for 4 number of simultaneous mixing requirements, a maximum of 8x8 chip size is sufficient to get optimum completion time by MLS.

### C. Benchmark Suites (BS I & BS III) and Hard Test Benches

The MLS algorithm is tested on Benchmark Suite-I (BS-I) , which comprises of 30 hard test benches and on most commonly used Benchmark Suite III (BS-III) [25]. We have also randomly generated 6 harder test cases [24] with more than 8 number of parallel mixing operations in a single stage. For such cases, it is difficult for the MLS/MMLS to find all chained mixing operations and thus increases completion time as well as length of Wash path ($W_{_path}$) which ultimately increases wash operation complexity as well as number of wash droplet requirement. MLS perform complete synthesis for all the 30 different test cases in BS-I well below 7 secs. For PCR and BS-III the detailed results are given in Table 6. Table 7 shows the detailed results for hard test cases (≥ 8 number of parallel mixing operations exist in a single stage) where combining SMT solver along with MLS-Shifts are giving better synthesis timing rather using MLS-shifts alone. For more than 8 number of simultaneous mixing operations, MLS unable to find chains for majority of the operations. Hence more number of Non-chain mixing operation increases path congestion problem on the chip which results in more number of "STALL" / Y-shift (-2.5% of -ve mixing) to come out of such deadlock situation. In such cases, from the deadlock point (Cell) applying SMT-solver along with MLS giving better completion time as given in Table 7.

## VIII. Conclusion

The proposed Module-Less-Synthesis (MLS_CP) method for cyberphysical DMFB provides faster bioassay results by eliminating the concept of dedicated virtual modules. Mixing operations are performed through any path, available on the microfluidic array and in accordance with the proposed shift-movements (patterns). According to the experimental results, MLC_CP leads to significant improvement in terms of assay synthesis time. Moreover, module-less synthesis is particularly important for more constrained synthesis problems as it can handle errors at an early stage of mixing operations and an improved error recovery technique can be incorporated which reduces overhead cost for the recovery operations. For more than 8 number of parallel mixing operations on a 13x13 chip (hard test benches), SMT solver is used to avoid congestion.

**Table 4:** Comparison of Synthesis completion times of IVD and PCR for different chip sizes

| Test Benches | Chip Size | Bio-Assay Completion Time (in Sec.) Module Based Synthesis Using ILP[12] | Using TS [12] | Proposed Module-less Synthesis |
|---|---|---|---|---|
| PCR | 8X9 | 9 | 8.9 | 5.75 |
| | 7X9 | 10 | 10 | 5.75 |
| | 7X8 | 14 | 13 | 6.25 |
| IVD | 8X9 | 13 | 12.5 | 9.0 |
| | 7X9 | 13 | 12.8 | 9.0 |
| | 7X8 | 14 | 13.7 | 9.5 |

**Table 5:** Chip Size Vs. Mixing Completion time of PCR (Stage I)

| *Chip Size* | *Time of completion of 4 parallel mixing operations (in sec)* |
|---|---|
| 4*X*4 | **3.263** |
| 5*X*5 | **3.075** |
| 6*X*6 | **1.750** |
| 7*X*7 | **1.512** |
| 8*X*8 | **1.375** |
| 12x12 | **1.319** |
| 16x16 | **1.301** |

**Table 6:** Synthesis Completion time by MLS for PCR and Benchmark Suite III and their respective Wash Path ($W_{_path}$) / Mixing Path ($M_{path}$) ratio

| **Test Benches (Chip- Size)** | **# of Mixing Operations (stage wise)** | **# of Chained Mixing Operations** | **Synthesis Time** | **($W_{_path}$ / $M_{_path}$)** |
|---|---|---|---|---|
| PCR (8x9) | 4 – 2 – 1 | 7 | 5.75 | |
| In_vitro I (16 x 16) | 6 | 6 | 3.1 | |
| In_vitroII (14 x 14) | 4 - 3 -2 | 9 | 6.1 | |
| Protein_I( 21 x 21) | 1-2-4-8-8-8-8 | 35 | 18.9 | |
| Protein_II(13x13) | (1-2-8-11-5-4) | 26 | 27.3 | |

**Table 7:** Comparison of Bioassay completion time for BS-III test cases (≥8 nos. of parallel mixing Operations) and Hard Test benches for SMT-solver disabled and SMT-solver enabled

| Test Benches | *SIZE* | *Max. # of parallel Mixing Operation* | *Completion time by MLS(SMT disabled)* | *COMPLETION TIME of MLS (SMT-Solver enabled)* |
|---|---|---|---|---|
| Protein-I | 21x21 | 8 | 18.9 | 13.2 |
| Protein-II | 13x13 | 11 | 27.3 | 14.1 |
| Hard Test1 | 24x24 | 14 | 22.6 | 16.7 |
| Hard Test2 | 16x16 | 14 | 28.9 | 21.4 |
| Hard Test3 | 13x13 | 12 | 26.4 | 18.4 |
| Hard Test4 | 12x12 | 12 | 29.0 | 21.7 |
| Hard Test5 | 12x12 | 10 | 24.4 | 17.9 |
| Hard Test6 | 12x12 | 09 | 21.5 | 17.3 |